\documentclass[12pt]{article}
\usepackage{amsmath,amssymb}    
\usepackage{amsfonts}
\usepackage{enumerate}
\usepackage{ytableau}
\usepackage{hyperref}
\usepackage{bm}
\usepackage{tikz}
\usepackage{blkarray}  
\begin{document}
\def\be{\begin{equation}}
\def\bea{\begin{eqnarray}}
\def\ee{\end{equation}}
\def\eea{\end{eqnarray}}
\def\d{\partial}
\def\eps{\varepsilon}
\def\la{\lambda}
\def\b{\bar}
\def\nn{\nonumber \\}
\def\p{\partial}
\def\t{\tilde}
\def\h{{1\over 2}}
\def\cp{\mathbb{CP}^1}
\makeatletter
\def\blfootnote{\xdef\@thefnmark{}\@footnotetext}  
\makeatother

\begin{center}
	{\LARGE  Comments on $\lambda$--deformed models \vspace{0.15cm} from 4D Chern-Simons theory}
	\\
	\vspace{18mm}
		{\bf Jia Tian}
	\vspace{14mm}
	
		Center for High Energy Physics (CHEP), \\Peking University,\\ No.5 Yiheyuan Rd, Beijing 100871, P.~R.~China
	
	\vspace{10mm}
	\blfootnote{email: wukongjiaozi@pku.edu.cn}
	
\end{center}
\begin{abstract}
	We study the $\lambda$--deformation of symmetric coset models from the viewpoint of a four dimensional Chern-Simons theory \cite{CY3}. In addition, by applying the ``dual'' boundary conditions of the ones used in the construction $\eta$--deformed PCM  in the trigonometric description \cite{TRIGYB} we construct a $\lambda$--deformation type model.
\end{abstract}

\tableofcontents

\baselineskip 18pt

\newpage

\section{Introduction}
A two dimensional classical integrable field theory is defined by the Lax connection. However there has been no systematic ways to derive the Lax connection given a integrable field theory or more bravely to classify integrable field theories.  Recently Costello and Yamazaki (CY) \cite{CY3} introduced a new point of view on two dimensional (2D) integrable field theories from a 4 dimensional (4D) Chern-Simons theory with a meromorphic one-form $\omega$ continuing their works on integrable lattice models \cite{CY12}. In this new approach, the 2D Lax connection is directly related to the 4D gauge field such that the flatness condition of the Lax connection is insured by the equation of motion of the gauge field. More significantly it provides a systematic way to construct 2D integrable field theories by specifying  a meromorphic one-form and boundary conditions of the gauge fields. Many interesting 2D integrable field theories have been realized in this approach \footnote{Interestingly, this one-form can be identified with a twist function which plays a crucial role in another new approach for constructing integrable field theories based on affine Gaudin model \cite{Gaudin1}. The relation between these two approaches is discussed in \cite{Gaudin2}.}\cite{Uni, COUPLED,HOLOLAMBDA,TRIGYB,NEWYB,STRING} including the Yang-Baxter deformation \cite{BosonicYangBaxter,DelducEta, Yoshida} and the $\lambda$--deformation \cite{Lambda,LambdaCoset}.  A very natural question is whether the construction can be generalized to include all the known classical integrable field theories. Hopefully by constructing enough examples we can understand the structure of integrability better and eventually classify integrable field theories.

In this paper our main aim is to generalize the result of $\lambda$--deformation to $\lambda$--deformed coset models \cite{LambdaCoset}. The strategy which is suggested in \cite{CY3} is to add cuts on the Riemann surface which the 4D  Chern-Simons theory depends on. Then we can impose a involution transformation when the gauge fields across the cut. On the end hand by passing to a double cover space the deck transformation can induce another algebra involution. Requiring the gauge fields to be invariant under the combined involution leads to a constraint on the gauge fields. Solving this constraint will restrict the gauge fields to take values in the coset space.
 
Another purpose of this paper is to construct the $\lambda$--deformation analogue of the Yang-Baxter model with the trigonometric description which is studied in \cite{TRIGYB}. Yang-Baxter model or more precisely the $\eta$-deformed model  admits two equivalent descriptions which will correspond to two different choices of holomorphic one-form and boundary conditions in the CY's approach. In \cite{TRIGYB}, the author revisited the CY's construction in the trigonometric description and found a new type $\eta$--deformation. It is well known that $\eta$--deformation is related the $\lambda$--deformation through the Poisson-Lie-T-duality \cite{PoissonLie}. Therefore it is natural to study the trigonometric description of  $\lambda$--deformation.

The paper has the following organization. In section \ref{Review}, we present a brief review of the CY 4D Chern-Simons theory and its relation to 2D integrable field theories. In section \ref{Trig} we construct the $\eta$--deformation and $\lambda$--deformation in the trigonometric description. In section \ref{Coset}, after implementing the construction of symmetric coset models following the suggestions given \cite{CY3}, we apply a similar strategy to construct $\eta$--deformation and $\lambda$--deformation of symmetric coset models.

\section{CY 4D Chern-Simons theory}
\label{Review}
\renewcommand{\theequation}{2.\arabic{equation}}
\setcounter{equation}{0}

In this section we briefly review the derivation of 2D integrable sigma models from CY 4D Chern-Simons theory approach and comment on the construction of the WZW model in \cite{CY3}.

\subsection{4D Chern-Simons theory}
We will choose the simplest set-up: the gauge group $G$\footnote{In general the group is complexified and one needs to impose reality condition during the construction but for our purpose considering real Lie algebra is enough} is a semi-simple Lie group and the corresponding Lie algebra is denoted by $\mathfrak{g}$ on which there exists a non-degenerate symmetric bilinear form $\langle,\rangle$. The 4D gauge field $A$ is defined on $\mathbb{R}^{2}\times \mathbb{CP}^1$. The action of the four-dimensional theory reads
\bea \label{4Daction}
S[A]=\frac{i}{4\pi} \int_{\mathbb{R}^{2} \times \mathbb{CP}^1} \omega \wedge CS(A),
\eea 
where 
\bea 
&&A=A_\sigma d\sigma+A_\tau d\tau+A_{\bar{z}}d\bar{z},\quad \omega=\omega(z)dz,\\
&&CS(A)=\langle A,dA+\frac{2}{3}A\wedge A\rangle.
\eea 
Here $\omega$ which is a meromorphic one-form on $\cp$ plays the central role in the construction. The positions of poles of $\omega$ are treated as boundaries of  $\cp$ and they are places where the resulted 2D integrable field theories live on. At the positions of zeros of $\omega$ we need to insert defect operators which describe the pole structures of the gauge fields such that their propagators are well defined.

Varying the action \eqref{4Daction} with respect to the gauge field $A$ gives
\bea 
\delta S[A]=\frac{i}{2\pi}\int_{\mathbb{R}^{2}\times \mathbb{CP}^1} \omega \wedge \langle \delta A,F\rangle +\frac{i}{4\pi}\int_{\mathbb{R}^{2}\times \mathbb{CP}^1} d\omega \wedge \langle A,\delta A\rangle,
\eea 
which leads to bulk equation of motion
\bea \label{BulkEOM}
\omega\wedge F(A)=0,\quad F(A)=dA+A\wedge A=0,
\eea 
and the boundary equation of motion
\bea \label{BEOM}
d\omega\wedge \langle A,\delta A\rangle=0.
\eea 
It's more useful to rewrite \eqref{BEOM} in terms of coordinates. Following the notation of \cite{Uni} let $\Sigma$ be the set of poles of $\omega(z) $ and $\xi_x$ be a local holomorphic coordinate around $x \in \Sigma$. Then the boundary condition \eqref{BEOM} can be rewritten as
\bea 
\sum_{x\in \Sigma}\sum_{p=0}^{m_x-1}(\text{Res}_x~ \xi^p_x \omega)\epsilon_{ij}\frac{1}{p!} \p^p_{\xi_x}\langle A_i, \delta A_j \rangle |_x=0,
\eea 
where $i,j= \sigma,\tau$ are the coordinates of $\mathbb{R}^{2}$ \footnote{We will identify the $\mathbb{R}^{2}$ at the all the positions of poles.}.
If we express $\omega$ as
\bea 
\omega(z)=\sum_{x}\sum_{p=0}^{m_x-1}\frac{l^x_p}{(z-x)^{p+1}}-l_\infty,
\eea 
then the boundary condition \eqref{BEOM} is 
\bea \label{Boundary2}
\sum_{x\in \Sigma}\sum_{p=0}^{m_x-1} \frac{l_p^x}{p!} \p^p_{\xi_x}\epsilon_{ij}\langle A_i, \delta A_j \rangle |_x=0.
\eea 
Note that the pole at infinity is also included in the expression above since $\xi_{\infty}=z^{-1}$.    
Introducing the light-cone coordinates $\sigma_{\pm}=\frac{1}{2}(\tau\pm \sigma)$ and we get
\bea 
\sum_{x\in \Sigma}\sum_{p=0}^{m_x-1}\frac{l_p^x}{p!}\p^p_{\xi_x}(\langle A_+,\delta A_-\rangle-\langle A_-,\delta A_+\rangle)|_x=0.
\eea 

\subsection{Lax connection and 2D action}
From the point of view of CY 4D Chern-Simons theory the Lax connection of the 2D integrable field theory is the fundamental object which is related to the 4D gauge field through a gauge transformation
\bea \label{ConnectionLax}
A=-d\hat{g} \hat{g}^{-1} +\hat{g}L\hat{g}^{-1},
\eea 
for some regular $\hat{g}:\mathbb{R}^{1,1}\times \cp \rightarrow G$ such that $L_{\bar{z}}=0$ or equivalently $A_{\bar{z}}=-\p_{\bar{z}} \hat{g}\hat{g}^{-1}$. In terms of Lax connection the bulk equation of motion \eqref{BulkEOM} takes a form of
\bea 
[\p_++L_+,\p_-+L_-]=0,\quad \omega \wedge \p_{\bar{z}}L(z,\tau,\sigma)=0.
\eea 
The first identity is the flatness condition of Lax connection. The second identity implies the positions of poles of $L$ coincide the positions of zeros of $\omega$. This fact guides us to make the ansatz of the Lax connection to solve the boundary conditions of the gauge fields. The field $\hat{g}$ will become to the field  of the 2D integrable field theory later on. To localize the four dimensional field theory to a two dimensional the field $g_x$, the 4D field $\hat{g}$ has to satisfy the archipelago conditions introduced in \cite{Uni} which we will not get into details.  Substituting \eqref{ConnectionLax} into the 4D action \eqref{4Daction} gives the final 2D action \cite{Uni}
\bea \label{Action}
S[\{g_x\}_{x\in \Sigma}]=\frac{1}{2} \sum_{x\in\Sigma} \int_{\mathbb{R}^{2}} \langle \text{Res}_x \omega \wedge L,g_x^{-1}dg_x\rangle-\frac{1}{2}\sum_{x\in \Sigma}(\text{Res}_x \omega) I_{WZ}[g_x].
\eea 
In this paper we will ignore the topological terms which can be restored easily when it is necessary. By choosing the gauge $A_{\bar{z}}=-\p_{\bar{z}} \hat{g}\hat{g}^{-1}$, the gauge symmetry has not been fully fixed. The residue gauge symmetry transfer to the gauge symmetry which is denoted by $H$ of $g_x$ . Therefore  if the gauge field $A$ at $x$ does not vanish we should remove this gauge redundancy $g_x\sim u_xg_x , u_x\in H, x\in \Sigma$ from  $g_x$. Besides that there is also a overall gauge transformation $g_x\rightarrow g_x h,  h\in G$ which does not modify the Lax connection.
\subsection{Boundary conditions}
Before constructing the 2D integrable field theories let us make some general comments about the boundary conditions \eqref{Boundary2}. Because the gauge fields are regular at those sites in $\Sigma$, so they admit a Taylor expansion with respect to $z$.  Let us focus on a generic pole $x$, then the gauge fields are expanded as
\bea 
A\sim \sum_p A_{[p]}^x \xi^p_x.
\eea 
We find that the boundary condition has no constraints on the components 
\bea\label{Blarge}
A_{[p]}^x \text{ is arbitrary},\quad p\geq m_x,
\eea 
where $m_x$ is the order of the pole. For the remaining components the boundary condition requires
\bea \label{Bsmall}
\sum_{x\in \Sigma}\sum_p l_{p+r}^x( \langle A^x_{+,[p]},\delta A^x_{+,[r]} \rangle -\langle A^x_{-,[p]},\delta A^x_{-,[r]}\rangle _{})=0 ,\quad p<m_\alpha.
\eea 
Since $\delta A_{+,[p]}$ and $\delta A_{-,[p]}$ are independent so these two terms should vanish separately
\bea 
\sum_{x\in \Sigma}\sum_p l_{p+r}^x \langle A^x_{+,[p]},\delta A^x_{+,[r]} \rangle =\sum_{x\in \Sigma}\sum_p l_{p+r}^x\langle A^x_{-,[p]},\delta A^x_{-,[r]}\rangle _{x}=0
\eea 
One possibility is 
\bea \label{Possibility1}
\sum_p l_{p+r}A^x_{+,[p]}=\sum_p l_{p+r} A^x_{-,[p]}=0,
\eea 
where the number of equations is same as the number of variables so generically we have the trivial solution which leads to the Dirichlet boundary condition.
Another possibility is that we can require
\bea 
\langle A^x_{[p]},\delta A^x_{[r]}\rangle =0
\eea 
by restricting $A^x$ take values only in the lagrangian subalgebra of $\mathfrak{g}$.
Particularly for the simple poles, the boundary condition is 
\bea 
\sum_{x\in \Sigma}l_0^x( \langle A_{}(x),\delta A_{}(x)\rangle=0,
\eea 
if some coefficients $l_0^x$ have the same magnitude we can group  them together to get a multiple-copied algebra $\mathfrak{g}^m=\mathfrak{g}\oplus \dots \oplus  \mathfrak{g}$, then the grouped gauge fields should take values in the lagrangian subalgebra of $\mathfrak{g}^m$. For example when $\omega$ has two simple poles $x_\pm$ such that $\text{Res}_{x_+}\omega=-\text{Res}_{x_-}\omega$ we can consider the direct sum $(A_{x_+},A_{x_-})\in \mathfrak{g}\oplus \mathfrak{g}$. The algebra $\mathfrak{g}\oplus \mathfrak{g}$ can be extended to a Manin triple $(\mathfrak{g}\oplus \mathfrak{g},\mathfrak{g}^R,\mathfrak{g}^\delta)$ defined as:
\bea \mathfrak{g}_R=\{(R-1)x,(R+1)x|x\in\mathfrak{g}\},\quad \mathfrak{g}^\delta=\{x, x|x\in\mathfrak{g}\} ,
\eea 
where the $R$ matrix satisfies the modified classical Yang-Baxter equation
\bea 
[Rx,Ry]-R([Rx,y]+[x,Ry])=-[x,y].
\eea 
With the choice of the bilinear form  
\bea 
\langle\langle (x,y),(x',y')\rangle\equiv\langle x,x'\rangle-\langle y,y'\rangle,
\eea 
both $\mathfrak{g}^\delta$ and $\mathfrak{g}_R$ are isotropic and lagrangian. Choosing one of them will lead to a 2D integrable field theory. The resulted two 2D integrable field theories are expected to be Poisson-Lie T-dual to each other \cite{PoissonLie}. One such example is the duality between $\eta$--deformation and $\lambda$--deformation.
\subsection{Comments on WZW model}
The simplest holomorphic 1-form on $\cp$ is
\bea \label{WZWForm}
\omega=dz/z.
\eea 
It has simple poles at  $0$ and $\infty$ but no zeros. The boundary conditions are so called  chiral Dirichlet \cite{CY3}:
\bea\label{BWZW} 
A_+|_0=0,\quad A_-|_\infty=0.
\eea
Before imposing the gauge symmetries the field contents are
\bea 
g_0=g\in G,\quad g_\infty=\tilde{g}\in G.
\eea 
Since there is no zeros in the one-form  we can parameterize the Lax connection as
\bea \label{AWZW}
L=L_+ d\sigma_++L_-d\sigma_-,
\eea 
where $L_\pm$ are regular functions on $\mathbb{R}^{1,1}\times \cp$.
Substituting the ansatz \eqref{AWZW} into the boundary conditions \eqref{BWZW} gives
\bea 
j_+=L_+,\quad \tilde{j}_-=L_-.
\eea 
Therefore the 2D action is 
\bea 
S=\frac{1}{2}\int ( \langle j_+,j_--\tilde{j}_-\rangle -\langle j_+-\tilde{j}_+,\tilde{j}_-\rangle)d\sigma_+\wedge d\sigma_- =\frac{1}{2}\int \langle j_+,\tilde{j}_-\rangle d\sigma_+\wedge d\sigma_-,
\eea 
where in the second equality we have used the residue gauge symmetry to fix
\bea 
j_-=\tilde{j}_+=0.
\eea 
However there is also an overall gauge symmetry which can be used to set $\tilde{g}=1$. This suggests that the resulting 2D theory should be the trivial one instead of WZW. The better way to describe WZW is from the limit of PCM with WZ term \cite{CY3}. 
\section{Deformed sigma models}
\label{Trig}
\renewcommand{\theequation}{3.\arabic{equation}}
\setcounter{equation}{0}

We are mostly interested in constructing $\lambda$--deformed coset model since the $\lambda$--deformed group model has been derived in \cite{Uni}. In \cite{TRIGYB} the author constructed the Yang-Baxter model from a trigonometric description in contrast to the rational one considered in \cite{Uni}.  The two equivalent descriptions are originated from the left-right duality of Yang-Baxter model \cite{LeftRight}. This section is dedicated to construct the $\lambda$-deformed model dual.
\subsection{Trigonometric description of $\eta$-deformation}
The meromorphic one-form $\omega$ in the trigonometric description is given by \footnote{In this section we mostly follow the notation in \cite{TRIGYB} but consider the real algebra case.}
\bea 
\omega=\frac{\sinh(\alpha-z)\sinh(\alpha+z)}{\sinh\alpha \cosh\alpha \sinh^2 z}dz
\eea 
Transferring from the cylinder to the plane via the map 
\bea 
w=\exp z,
\eea 
we can obtain a rational one-form:
\bea 
\omega=\frac{4(e^{2\alpha}-w^2)(e^{2\alpha}w^2-1)}{(e^{4\alpha}-1)w (w^2-1)^2}d\omega.
\eea 
The set of simple poles $\mathfrak{p}_1$ and double poles $\mathfrak{p}_2$ are
\bea 
\mathfrak{p}_1=\{0,\infty\},\quad \mathfrak{p}_2=\{1,-1\}.
\eea 
At double poles $w=\pm 1$ we impose the  Dirichlet boundary conditions:
\bea\label{BTrigYB1}
A|_1=A|_{-1}=0.
\eea 
The boundary conditions are different from those used in \cite{TRIGYB} but they are equivalent. At simple poles $w=0,\infty$ the residues are opposite so we can choose the boundary condition
\bea \label{BTrigYB2}
(A|_0,A|_\infty)\in \mathfrak{g}^R.
\eea 
Using the gauge symmetry we can first set the fields  to be
\bea 
g_{-1}=g_-\in G,\quad g_{1}=g_+\in G,\quad (g_0,g_\infty)=(g,g),
\eea 
then using the overall gauge symmetry to fix $g=1$.  The relation \eqref{ConnectionLax} gives
\bea 
&&A_0=L|_{0},\quad A_
\infty=L|_{\infty},\nn
&&A_1=-dg_+ g^{-1}_+ +g_+ L_{1} g_+^{-1},\quad A_{-1}=-dg_- g^{-1}_- +g_- L_{-1} g_-^{-1}.
\eea 
Considering that there are four zeros in $\omega(w)$ and to avoid the appearance of double poles in the flatness condition we take the ansatz of the Lax connection $L$ to be 
\bea \label{TrigLax}
L_+=\frac{V_+ \omega+V_+'}{e^{2\alpha}\omega^2-1}+U_+,\quad L_-=\frac{V_- \omega+V_-'}{\omega^2-e^{2\alpha}}+U_-,
\eea 
where $V_\pm,V_\pm'$ and $U_\pm$ are regular functions.
Substituting the ansatz into the boundary conditions \eqref{BTrigYB1} and \eqref{BTrigYB2} one can get
\bea 
&&V_\pm=\pm(e^{2\alpha}-1)\frac{j_\pm^\oplus-j_\pm^\ominus}{2},\quad j^\oplus=g^{-1}_+ dg_+,\quad j^\ominus=g^{-1}_- dg_-,\nn
&&V_+'=\frac{j_+^\oplus+j_+^\ominus}{R+\lambda_\alpha},\quad V_-'=e^{2\alpha}\frac{j_-^\oplus+j_-^\ominus}{R-\lambda_\alpha},\quad \lambda_\alpha=\frac{e^{2\alpha}+1}{e^{2\alpha}-1},\nn
&&U_\pm=\frac{j_\pm^\oplus+j_\pm^\ominus}{2}\mp \frac{V_\pm'}{e^{2\alpha}-1},
\eea 
where the $\mathfrak{g}$ valued currents are defined as $j^{\oplus}=g^{-1}_+ dg_+$ and $j^{\ominus}=g^{-1}_- dg_-$.
To derive the 2D action we need to evaluate $\text{Res}_{\pm 1}\omega \wedge L$. One should be careful that these are residues for the double poles so they pick the coefficient of $(\omega \mp 1)$ of  $L$ in the Taylor series expansion. The results are
\bea 
&&\text{Res}_1 (\omega(w) L_+)=-\frac{V_+}{e^{2\alpha}-1}-\frac{2e^{2\alpha}V_+'}{e^{4\alpha}-1},\quad \text{Res}_{1} (\omega(w) L_-)=-\frac{V_-}{e^{2\alpha}-1}-\frac{2V_-'}{e^{4\alpha}-1},\nn
&&\text{Res}_{-1} (\omega(w) L_+)=\frac{V_+}{e^{2\alpha}-1}-\frac{2e^{2\alpha}V_+'}{e^{4\alpha}-1},\quad \text{Res}_{1} (\omega(w) L_-)=\frac{V_-}{e^{2\alpha}-1}-\frac{2V_-'}{e^{4\alpha}-1},\nn
\eea 
which lead to the 2D action
\bea \label{TrigYBaction}
-\frac{1}{2} \int d\sigma_+\wedge d\sigma_- \left[\langle j_+^\oplus-j_+^\ominus,j_-^\oplus-j_-^\ominus\rangle+\langle j_+^\oplus+j_+^\ominus,\frac{(1-\eta^2)}{1-\eta R}(j_-^\oplus+j_-^\ominus)\rangle \right],
\eea 
where we have introduced the deformation parameter $\eta=\frac{1}{\lambda_\alpha}$.

In general $j^\oplus$ and $j^\ominus$ are independent then the action actually does not describe a Yang-Baxter model. While the authors in \cite{TRIGYB} obtained the (generalized) Yang-Baxter model by imposing some relations between $j^\oplus$ and $j^\ominus$. They observed that the action is invariant under swapping $j^\oplus$ and $j^\ominus$ so these two currents should be related by a involution. This argument is kind of ad hoc. Perhaps a better way to view this is to notice that the parameter in the trigonometric description is related to the parameter in the rational description via a Mobius transformation \cite{LeftRight}
\bea 
\frac{1}{w^2}=\frac{z-\eta}{z+\eta}
\eea 
So we should think of that the $w$ space is a double cover of $z$ plane. Since $w=\pm 1$ have the same preimage we should identify $j^\oplus$ and $j^\ominus$. Alternatively we can think of that the cut is a topological domain wall \cite{CY3} and when we cross the wall, we can apply a automorphism $\rho$ of the algebra $\mathfrak{g}$ as
\bea 
j^\ominus=\rho (j^\oplus),\quad \rho^2=1.
\eea

\subsection{Trigonometric description of $\lambda$ deformation}
Now we consider the Poisson-Lie-T-dual of the Yang-Baxter model in the trigonometric description by choosing  the boundary conditions at the simple poles to be
\bea \label{BTrigYB3}
(A|_0,A|_\infty)\in \mathfrak{g}^\delta.
\eea 
Using this gauge symmetry and the overall gauge symmetry we can fix the fields to be 
\bea
g_{-1}=g_{\infty}=1,\quad g_1=g\in G,\quad g_0=\tilde{g}\in G.
\eea 
In this case we will not impose the double cover condition anymore since the gauge fixing breaks the swapping symmetry. The relation \eqref{ConnectionLax} implies
\bea \label{triglambdaboundary}
&&A_0=-d\tilde{g} \tilde{g}^{-1} +Ad_{\tilde{g}} L|_0,\quad A_1=-dg g^{-1}+Ad_g L|_1,\nn
&&A_{-1}=L|_{-1},\quad A_{\infty}=L|_{\infty}.
\eea 
Substituting the same ansatz \eqref{TrigLax} into the boundary conditions \eqref{BTrigYB1} and \eqref{BTrigYB3} gives
\bea 
&&j_\pm=\pm \frac{V_\pm+V_\pm '}{e^{2\alpha}-1}+U_\pm,\quad \pm\frac{V_\pm '-V_\pm}{e^{2\alpha}-1}+U_\pm=0,\nn
&& U_+=-\p_+\tilde{g}\tilde{g}^{-1}+\tilde{D}(-V_+'+U_+),\nn
&&U_-=-\p_-\tilde{g}\tilde{g}^{-1}+\tilde{D}(-e^{-2\alpha}V_-'+U_-).
\eea 
where we have defined the operator $\tilde{D}=Ad_{\tilde{g}^{}}$ .These equations can be solved by
\bea 
&&V_\pm=\pm (e^{2\alpha}-1)\frac{j_\pm}{2},\nn
&&V_\pm '=\pm\frac{e^{2\alpha}-1}{e^{\pm 2\alpha}-\tilde{D}^T}\left[(1-\tilde{D}^T)\frac{j_\pm}{2}-\tilde{j}_{\pm}\right],\nn
&&U_\pm=\frac{j_\pm}{2}\mp \frac{V_\pm '}{e^{2\alpha}-1}.
\eea 
To derive the 2D action  we need to evaluate the residues $\text{Res}_{1}\omega \wedge L$ and $\text{Res}_{0}\omega \wedge L$ which are given by
\bea 
&&\text{Res}_1 (\omega(w) L_+)=-\frac{V_+}{e^{2\alpha}-1}-\frac{2e^{2\alpha}V_+'}{e^{4\alpha}-1},\quad \text{Res}_{1} (\omega(w) L_-)=-\frac{V_-}{e^{2\alpha}-1}-\frac{2V_-'}{e^{4\alpha}-1},\nn
&&\text{Res}_0 (\omega(w) L_+)=\frac{4e^{2\alpha}}{e^{4\alpha}-1}(V_+'-U_+),\quad \text{Res}_0 (\omega(w) L_-)=\frac{4e^{2\alpha}}{e^{4\alpha}-1}(e^{-2\alpha}V_-'-U_-).\nn
\eea 
The resulting 2D action reads
\bea \label{triglamnbdaaction}
&&S[j,\tilde{j}]=\frac{2e^{2\alpha}}{e^{4\alpha}-1}\int \left[(\langle \tilde{j}_+,\tilde{j}_- \rangle +2 \langle \frac{\tilde{D}}{e^{-2\alpha}-\tilde{D}}\tilde{j}_+,\tilde{j}_-\rangle) d\sigma_+\wedge d\sigma_-)\right]d\sigma_+\wedge d\sigma_-\nn
&&\qquad\quad  -\frac{1}{2} \int \left[ \langle j_+,j_-\rangle+\frac{2}{e^{2\alpha}+1}\langle \frac{1-\tilde{D}}{e^{-2\alpha}-\tilde{D}}j_+,j_-\rangle  \right]d\sigma_+\wedge d\sigma_-\nn
&&\qquad\quad  +\frac{2}{1+e^{2\alpha}}\int \left[\langle \frac{1}{e^{-2\alpha}-\tilde{D}}j_+,\tilde{j}_-\rangle+\langle \tilde{j}_+,\frac{e^{2\alpha}}{e^{2\alpha}-\tilde{D}}j_-\rangle \right]d\sigma_+\wedge d\sigma_-.
\eea 
The first line in the action really describes the $\lambda$--deformed model up to a overall factor if we identify $e^{2\alpha}\equiv \lambda$.  The whole action describes a $\lambda$--deformed model coupled with another $\lambda$-deformed like sigma model. This coupled model may relate to the coupled $\lambda$--models constructed  in \cite{COUPLED}.

\section{Deformed coset models}
\label{Coset}
\renewcommand{\theequation}{4.\arabic{equation}}
\setcounter{equation}{0}
In the original CY's paper \cite{CY3}, it is proposed that coset models can be constructed by introducing a cut in the Riemann surface as we discussed in last section. Recently in \cite{NEWYB} the homogeneous Yang-Baxter deformed coset model is constructed in a different way. In this section, we first show the details of the construction of symmetric coset  models following CY's original suggestion then extend the construction to  $\eta$-- and $\lambda$-- deformed coset models.
\subsection{Symmetric coset model}
We consider a coset $G/H$ with corresponding Lie algebra $\mathfrak{g}$ of $G$ and $\mathfrak{h}$ of $H$. The coset is called symmetric if the Lie algebra $\mathfrak{g}$ admits a $Z_2$-grading:
\bea 
&&\mathfrak{g}=\mathfrak{h}\oplus \mathfrak{m},\quad [\mathfrak{h},\mathfrak{h}]\subset \mathfrak{h},\quad [\mathfrak{h},\mathfrak{m}]\subset \mathfrak{m},\quad [\mathfrak{m},\mathfrak{m}]\subset \mathfrak{h}.
\eea 
To construct the symmetric coset sigma model we put a cut at the interval $[-1-\lambda^2,1+\lambda^2]$ at the Riemann sphere $\cp$. Introducing a coordinate $u$ by implementing the Joukowsky transform:
\bea \label{JouTran}
\lambda^{-1}u+\lambda u^{-1}=z
\eea 
we can get a double cover with a deck transformation $u\rightarrow \lambda^2 u^{-1}$ of the z-plane. The one-from $dz$ pulls back to 
\bea 
\omega=\frac{(u-\lambda)(u+\lambda)}{\lambda u^2}du.
\eea 
On the double cover $u$ plane the set of simple zeros and double poles of the one-form are
\bea 
\mathfrak{z}=\{\pm\lambda\},\quad \mathfrak{p}_2=\{0,\infty\}.
\eea 
At the two boundaries $u=0$ and $u=\infty$ there are fields
\bea \label{coset1}
g_0=g\in G,\quad g_\infty=\tilde{g}\in G.
\eea 
Because $u=0$ and $u=\infty$  have the same preimage so they are related by the deck transformation we should have the identification
\bea \label{coset2}
g=\tilde{g}\quad \text{or}\quad \tilde{g}=\rho (g),
\eea 
where $\rho$ is a involution. The first choice will lead to a trivial theory so we will choose the second one with the $Z_2$ involution:
\bea \label{Z2}
\rho (\mathfrak{h})=\mathfrak{h},\quad \rho (\mathfrak{m})=-\mathfrak{m}.
\eea 
At the double poles we apply the Dirichlet boundary conditions:
\bea \label{BCoset}
A_0=A_\infty=0.
\eea 
The relation \eqref{ConnectionLax} gives
\bea \label{coset}
A_0=-dgg^{-1}+Ad_g L|_0,\quad A_\infty=-d\tilde{g}\tilde{g}^{-1}+Ad_{\tilde{g}} L|_\infty.
\eea 
Consider the zeros of the one-form are $\pm \lambda$ we assume that the Lax connection is given by
\bea\label{CosetAnsatz}
L_+=\frac{u+\lambda}{u-\lambda}V_++U_+,\quad L_-= \frac{u-\lambda}{u+\lambda}V_-+U_-.
\eea
Substituting the ansatz into the boundary condition \eqref{BCoset}, one can obtain 
\bea 
V_\pm=-\frac{1}{2}(j_\pm-\tilde{j}_\pm),\quad U_\pm=\frac{1}{2}(j_\pm+\tilde{j}_\pm).
\eea 
The residues of $\omega(u)L$ at $u=0,\infty$ are evaluated as
\bea 
&&\text{Res}_0(\omega(u)L)=2(V_+ d\sigma_+-V_-d\sigma_-),\\
&&\text{Res}_\infty(\omega(u)L)=-2(V_+ d\sigma_+-V_-d\sigma_-).
\eea 
Therefore the resulted  2D action \eqref{Action} is
\bea 
S[j,\tilde{j}]=-\int \langle j_+-\tilde{j}_+,j_--\tilde{j}_-\rangle d\sigma_+\wedge d\sigma_-.
\eea 
By employing the $Z_2$ involution \eqref{Z2} one can find
\bea 
&&S=-2 \int \langle j_+^{(1)},j_-^{(1)} \rangle d\sigma_+\wedge d\sigma_1,\nn
&& L_\pm=j_\pm ^{(0)}+(\frac{\lambda+u}{\lambda-u})^{\pm}j_\pm^{(1)},\quad j^{(0)}\in \mathfrak{h},~j^{(1)}\in \mathfrak{m}.
\eea 
These describe the standard symmetric coset model.
\subsubsection*{Alternative prescription}
The choice of field contents \eqref{coset1} and \eqref{coset2} seems to be artificial. Let us understand it in our general picture. Before imposing any gauge symmetry, the field contents should be $(g,\tilde{g})$. As usual, we can use the overall gauge symmetry to fix one of them to be identity, for example $(g,1)$. The involution condition of $g$ imposes the constraint
\bea \label{Cons}
g^{-1}=\rho(g).
\eea 
 Now the relation \eqref{ConnectionLax} gives
\bea 
A_\infty=L|_\infty,\quad A_0=-dg g^{-1}+g L|_0 g^{-1}.
\eea 
Using the same ansatz \eqref{CosetAnsatz} of the Lax connection and substituting them into the boundary conditions \eqref{BCoset}
one can solve
\bea 
U_\pm=0,\quad V_\pm=-j_\pm.
\eea
Evaluating the residue in \eqref{Action} we find that 
\bea \label{actioncoset}
S[j]\sim \int \langle j_+,j_-\rangle d\sigma_+\wedge d\sigma_-.
\eea 
However this is not the final action because there is a constraint \eqref{Cons}. We can solve the constraint by 
\bea\label{Solve}
&&g\equiv\rho(g')g'^{-1},\quad g'\in G, \nn
&&j=[\rho(g')g'^{-1}]^{-1}d[\rho(g')g'^{-1}]=g(\rho(j)-j)g^{-1}=-2g j^{(1)} g^{-1}
\eea
where in the last two equalities we have renamed $g'$ with $g$.
Therefore substituting \eqref{Solve} into \eqref{actioncoset} we end up with the standard action of the coset model
\bea 
S[j^{(1)}]\sim \int \langle j_+^{(1)},j_-^{(1)}\rangle d\sigma_+\wedge d\sigma_-.
\eea 
\subsection{$\eta$--deformed coset model}
Let us consider the one-form 
\bea 
\omega(u)=\frac{K}{2}\frac{du}{(u-\alpha^2)(u-\beta^2)},
\eea
which relates to $dz/z$ via some Mobius transformation. Introducing the coordinate $u=z^2$, the one-form pulls back to
\bea\label{CosetForm}
\omega(z)=\frac{K z dz}{(z-\alpha)(z-\beta)(z+\alpha)(z+\beta)}.
\eea 
This means the $z$ space is a double cover of $u$--plane with the deck transformation $z\rightarrow -z$. The one-from on z--plane has four simple poles at $\pm \alpha,\pm \beta$ with residues:
\bea 
\text{Res}_{-\alpha}\omega=\text{Res}_{\alpha}\omega=-\text{Res}_{-\beta}\omega=-\text{Res}_{\beta}\omega=\frac{K}{2}\frac{1}{\alpha^2-\beta^2}.
\eea 
Therefore we can impose the following boundary conditions:
\bea \label{BYBCoset}
(A|_{-\alpha},A_{-\beta})\in \mathfrak{g}^R,\quad (A|_{\alpha},A_{\beta})\in \mathfrak{g}^{\tilde{R}}
\eea 
The residue gauge symmetry at the boundaries can be imposed to fix fields to be
\bea 
g_{-\alpha}=g_{-\beta}=g\in G,\quad g_{\alpha}=g_{\beta}=\tilde{g}\in G.
\eea 
Because the one-form only has zeros at $z=0$ and $z=\infty$ the Lax connection takes a form of
\bea\label{LaxCoset}
&&L_+=\frac{z+\alpha}{2\alpha}V_+-\frac{z-\alpha}{2\alpha}U_+,\nn
&&L_-=\frac{z+\alpha}{2z}V_-+\frac{z-\alpha}{2z}U_-,
\eea 
where $V_\pm$ and $U_\pm$ are regular functions. Substituting the  ansatz of the Lax connection \eqref{LaxCoset} into the boundary conditions  \eqref{BYBCoset} gives the following equations:
\bea \label{Eqn}
&&-2j_+=(R_g-1)\frac{\alpha-\beta}{2\alpha}V_++[(R_g-1)\frac{\alpha+\beta}{2\alpha}-(R_g+1)]U_+,\nn
&&-2\tilde{j}_+=[(\tilde{R}_{\tilde{g}}-1)\frac{\alpha+\beta}{2\alpha}-(\tilde{R}_{\tilde{g}}+1)]V_++(\tilde{R}_{\tilde{g}}-1)\frac{\alpha-\beta}{2\alpha}U+,\nn
&&-2j_-=(R_g-1)\frac{\beta-\alpha}{2\beta}V_-+[(R_g-1)\frac{\alpha+\beta}{2\beta}-(R_g+1)]U_-,\nn
&&-2\tilde{j}_-=(\tilde{R}_{\tilde{g}}-1)\frac{\beta-\alpha}{2\beta}U_-+[(\tilde{R}_{\tilde{g}}-1)\frac{\alpha+\beta}{2\beta}-(\tilde{R}_{\tilde{g}}+1)]V_.
\eea 
Solving for $V_\pm$ and $U_\pm$ is cumbersome but straightforward however by evaluating the residues in \eqref{Action} one finds that 2D action only depends on a special linear combination of $U$ and $V$:
\bea 
&&S=\int \mathcal{L}~d\sigma_+\wedge d\sigma_-,\nn
&&\mathcal{L}=\frac{K}{4}\frac{1}{\alpha^2-\beta^2}[(\langle U_+,j_-\rangle-\langle j_+,U_-\rangle)+(\langle V_+,\tilde{j}_-\rangle-\langle \tilde{j}_+, V_-\rangle)\nn
&&\qquad\qquad -(\langle\frac{\alpha-\beta}{2\alpha}V_++\frac{\alpha+\beta}{2\alpha}U_+,j_-\rangle-\langle j_+,-\frac{\alpha-\beta}{2\beta} V_-+\frac{\alpha+\beta}{2\beta}U_-)\nn
&&\qquad\qquad -(\langle \frac{\alpha+\beta}{2\alpha}V_+-\frac{\beta-\alpha}{2\alpha}U_+,\tilde{j}_-\rangle-\langle \tilde{j}_+,\frac{\alpha+\beta}{2\beta}V_-+\frac{\beta-\alpha}{2\beta}U_-)]\nn
&&\quad =\frac{K}{4}\frac{1}{\alpha^2-\beta^2}[-\frac{\alpha-\beta}{2\alpha}\langle V_+-U_+,j_--\tilde{j}_-\rangle-\frac{\alpha-\beta}{2\beta}\langle j_+-\tilde{j}_+,V_--U_-\rangle]
\eea 
After some algebraic manipulation of \eqref{Eqn} we get
\bea 
&&V_+-U_+=\frac{2\alpha}{\alpha+\beta}\frac{\tilde{j}_+-j_+}{1+\eta R_g+\eta \tilde{R}_{\tilde{g}}},\nn
&&V_--U_-=\frac{2\beta}{\alpha+\beta}\frac{\tilde{j}_--j_-}{1-\eta R-\eta \tilde{R}_{\tilde{g}}},
\eea 
where we have introduced the new parameter
\bea 
\eta\equiv \frac{\alpha-\beta}{2(\alpha+\beta)}.
\eea 
Therefore the final 2D action is 
\bea 
S=\frac{K}{4}\frac{1}{(\alpha+\beta)^2}\int \langle j_+-\tilde{j}_+,\frac{1}{1-\eta R-\eta \tilde{R}_{\tilde{g}}} (j_--\tilde{j}_-)\rangle.
\eea 
We have not imposed the overall gauge symmetry. One can use it to set $\tilde{g}=1$, then the resulted action is the bi-Yang-Baxter model \cite{BYB}. Because our symmetric choice of positions of poles, there is only one deformation parameter $\eta$ instead of two as in \cite{Uni} . 

Alternatively one can impose the involution condition on the double cover space:
\bea 
\tilde{g}=\rho{(g)},\quad \tilde{j}=\rho({j}),
\eea 
if the algebra $\mathfrak{g}$ admits a $Z_2$-grading $\mathfrak{g}=\mathfrak{g}^{(1)}\oplus \mathfrak{g}^{(0)}$. 

Let $g=\exp(T^{(0)}+T^{({1})})$, then the Campbell-Baker-Hausdorff formula implies
\bea 
g B g^{-1}=B+[T^{(0)}+T^{({1})},B]+\frac{1}{2}[T^{(0)}+T^{({1})},[T^{(0)}+T^{({1})},B]]+\dots
\eea
Separating $B$ into two components according to the $Z_2$-grading one can get
\bea 
&&(g B g^{-1})^{(0)}=[T^{(0)},B^{(0)}]+[\text{even number of $T^{(1)}$},B^{(0)}]+[\text{odd number of $T^{(1)}$},B^{(1)}],\nn
&&\equiv D_{ab}B_b^{(0)}+D_{a\alpha}B_{\alpha}^{(1)}\\
&&(g B g^{-1})^{(1)}=[T^{(0)},B^{(1)}]+[\text{even number of $T^{(1)}$},B^{(1)}]+[\text{odd number of $T^{(1)}$},B^{(0)}],\nn
&&\equiv D_{\alpha b}B_b^{(0)}+D_{\alpha\beta}B_{\alpha}^{(1)}
\eea 
The $Z_2$ involution implies
\bea \label{Z2Ad}
&&(\tilde{g} B \tilde{g}^{-1})^{(0)}=D_{ab}B_b^{(0)}-D_{a\alpha}B_{\alpha}^{(1)},\nn
&&(\tilde{g} B \tilde{g}^{-1})^{(1)}=-D_{\alpha b}B_b^{(0)}+D_{\alpha\beta}B_{\alpha}^{(1)}.
\eea 
Separating the generators $T^A=(T^a,T^\alpha)$ of the group $G$ into $T^a$ and $T^\alpha$ corresponding to the subgroup $H$ and the coset $G/H$  respectively, the adjoined action $Ad_g=D$ and the $Z_2$ involution $\rho$ can be expressed as explicit matrices
\bea \label{Dab}
D_{AB}=\mbox{Tr}(T_A g T_B g^{-1})=\begin{Bmatrix}
	d_1&d_2\\d_3&d_4
\end{Bmatrix},\quad \rho=\begin{Bmatrix}
	1&0\\0&-1
\end{Bmatrix}.
\eea 
According to \eqref{Z2Ad} the $Z_2$ involution acts on $D$ as
\bea \label{Dab1}
\tilde{D}_{AB}=\mbox{Tr}(T_A \tilde{g} T_B \tilde{g}^{-1})=\begin{Bmatrix}
	d_1&-d_2\\-d_3&d_4
\end{Bmatrix}=\rho D\rho.
\eea 
Combining \eqref{Dab} and \eqref{Dab1} we obtain
\bea 
R_g+\tilde{R}_{\tilde{g}}=DRD^T+\rho{D} RD^T\rho=2P^{(0)}R_gP^{(0)}+2P^{(1)}R_gP^{(1)},
\eea 
where $P^{(0)}$ and $P^{(1)}$ is the grading-0 and grading-1 projector respectively. 
Then we find that in the double cover situation the 2D action is 
\bea 
S=\frac{K}{(\alpha+\beta)^2}\int \langle j_+^{(1)},P^{(1)}\frac{1}{1-2\eta R_gP^{(1)} } j_-^{(1)}\rangle,
\eea 
which describes the Yang-Baxter coset model. Differ from the similar construction of Yang-Baxter coset model in \cite{NEWYB}, in our case the matrix $R$ is the solution of the modified classical Yang-Baxter equation instead of the homogeneous one considered in \cite{NEWYB}.

\subsection{$\lambda$--deformed coset model}
Let us consider the same  one-form 
\bea
\omega(z)=\frac{K z dz}{(z-\alpha)(z-\beta)(z+\alpha)(z+\beta)}.
\eea 
but choose  the ``dual'' boundary conditions
\bea \label{BlambdaCoset}
(A|_{-\alpha},A_{-\beta})\in \mathfrak{g}^\delta,\quad (A|_{\alpha},A_{\beta})\in \mathfrak{g}^{\delta}
\eea
Taking the same ansatz \eqref {LaxCoset} and substituting them into the new boundary conditions \eqref{BlambdaCoset} give
\bea 
&&j_+=-D^TU_++\frac{\alpha-\beta}{2\alpha}V_++\frac{\alpha+\beta}{2\alpha}U_+,\nn
&&\tilde{j}_+=-\tilde{D}^T V_++\frac{\beta+\alpha}{2\alpha}+\frac{\alpha-\beta}{2\alpha}U_+,\nn
&&j_-=-\frac{\alpha-\beta}{2\beta}V_-+\frac{\alpha+\beta}{2\beta}U_--D^T U_-,\nn
&&\tilde{j}_-=\frac{\alpha+\beta}{2\beta}V_-+\frac{\beta-\alpha}{2\beta}U_--\tilde{D}^TV_-,
\eea 
where $U_\pm$ and $V_\pm$ are  regular functions. We are aiming to construct the coset model so we again separate all the quantities into two components according to the $Z_2$ grading of the algebra. Using \eqref{Dab} and \eqref{Dab1}, one can find
\bea \label{MN}
&&2j_\pm^{(0)}=(1-d_1^T)M_\pm^{(0)}+d_3^T N_\pm^{(1)},\nn
&&2j_\pm^{(1)}=-(\eta^{\pm}-d_4^T)N^{(1)}_\pm-d_2^T M^{(0)}_\pm,
\eea 
where we have defined convenient variables
\bea 
\eta=\beta/\alpha,\quad M_\pm=V_\pm+U_\pm,\quad N_\pm=V_\pm-U_\pm.
\eea 
Evaluating the residues at $z=\pm \beta$ in \eqref{Action} we find the 2D action is given by
\bea \label{actionlambda}
&&S=-\frac{K}{2}\frac{1}{\alpha^2-\beta^2}\int (\langle M_+^{(0)},j_-^0\rangle+\eta\langle -N_+^{(1)},j_-^{(1)}\rangle)-(\langle j_+^{(0)},M_-^{(0)}\rangle+\eta^{-1}\langle j_+^{(1)},-N^{(1)}\rangle)\nn
&&\quad =-\frac{K}{2}\frac{1}{\alpha^2-\beta^2}\int  \langle J_+,\Lambda j_-\rangle-\langle j_+,\Lambda^{-1}J_-\rangle,
\eea 
where
\bea 
J_\pm=M_\pm^{(0)}-N^{(1)}_\pm,\quad \Lambda=\begin{Bmatrix}
	1&0\\0&\eta
\end{Bmatrix}.
\eea 
From \eqref{MN}, one can solve $J_\pm$ as
\bea \label{J}
J_\pm=2P^{-1}j_\pm,\quad P=\begin{Bmatrix}
	1-d_1^T&-d_3^T\\-d_2^T&\eta^{\pm}-d_4^T
\end{Bmatrix}.
\eea 
Substituting \eqref{J} into \eqref{actionlambda} results
\bea 
S=\frac{K}{\alpha^2-\beta^2}\int(\langle j_+,j_-\rangle+2 \langle j_+,\frac{1}{\Lambda^{-1}-D^T}D^Tj_-\rangle
\eea 
which coincides with the action of the $\lambda$ coset model \cite{LambdaCoset} up to a overall factor which can be absorbed into $K$.

\section{Discussion}
In this paper, we have discussed (trigonometric) $\lambda$--deformed (coset) models from the viewpoint of CY 4D Chern-Simons theory. In the case of trigonometric $\lambda$--deformed model we find that the resulted 2D theory describes a coupled $\lambda$--deformed like sigma model. The situation is different from the trigonometric Yang-Baxter model where two coupled sigma models admit a swapping symmetry which can be used to remove half of degrees of freedom. The main result of the this paper is to realize the $\lambda$-deformed coset models from the 4D Chern-Simons theory by adding cuts in the Riemann surface. After introducing double cover space the involution condition appears naturally. A similar analysis is applicable for the $\lambda$--deformed $AdS_5\times S^5$ superstring which we will report in the future work.

It is interesting to extend the current analysis to other generalized $\lambda$--deformed models including the asymmetric $\lambda$--deformation \cite{ASY} and $\lambda$-Yang-Baxter models \cite{GenLambda1}. It would be also interesting to attempt to combine the CY's approach and the affine Gaudin model approach to  develop a more powerful tool.

\section*{Acknowledgments}
I would like to thank the Tohoku University for the hospitality during the 14th Kavli Asian Winter School on Strings, Particles and Cosmology. It is my pleasure to attend M. Yamazaki's lectures at the Winter School. I also want to thank Jue Hou, Han Liu, Yijun He and Jun-Bao Wu for useful discussions.

\appendix

\end{document}